%% file: main.tex
\documentclass[runningheads]{llncs}
\usepackage{graphicx}
\usepackage{todonotes} 

 \usepackage{comment}
\usepackage{array}

 \usepackage[colorlinks=true,
        linkcolor=black,
        citecolor=black,
        filecolor=black,
        urlcolor=black,
        bookmarks=true,
        bookmarksopen=true,
        bookmarksopenlevel=3,
        plainpages=false,
        pdfpagelabels=true]{hyperref}

\input{aux-commands}
\begin{document}
\title{Neurodiversity in Agile Teams:\\ Obstacles and Inclusion Barriers}
\titlerunning{Neurodiversity in Agile Teams}
%
\author{Lars Struck\inst{1}\orcidID{0009-0006-3804-5273}\and
Christian Veenaas\inst{1}\orcidID{0009-0001-2417-6057}\and
Robert Wiedekind\inst{1}\orcidID{0009-0004-5021-8531}\and
Joshua Riechmann\inst{1}\orcidID{0009-0000-1244-9129}\and
Maria Rauschenberger\inst{2}\orcidID{0000-0001-5722-576X}   \and
Philipp Diebold\inst{4,5}\orcidID{0000-0002-3910-7898} \and
Simone Dogu\inst{3}\orcidID{}\and
Michael Neumann\inst{1}\orcidID{0000-0002-4220-9641}
}

\authorrunning{L. Struck et al.}
%
\institute{University of Applied Sciences and Arts Hannover,\\ Hannover, Germany 
\email{michael.neumann@hs-hannover.de}\\
\and
University of Applied Sciences Emden/Leer,\\  Emden/Leer, Germany
\email{maria.rauschenberger@hs-emden-leer.de}\\
\and
DB Systel GmbH, Frankfurt/Main, Germany\\
\email{simone.dogu@deutschebahn.com}
\and
Bagilstein GmbH,  Im Niedergarten 10,  55124 Mainz , Germany\\
\email{philipp.diebold@bagilstein.de}
\and
IU International University, Juri-Gagarin-Ring 152, 99084 Erfurt, Germany\\
\email{philipp.diebold@iu.org}
}
\maketitle              
\begin{abstract}
\textit{Context:} Neurodiversity is increasingly recognized as a valuable dimension of workplace diversity. However, in agile software development teams, the interplay between teamwork practices and the inclusion of neurodivergent employees remains underexplored.
\textit{Objective:} The study aims to explore how teamwork quality in agile software development is currently practiced and discussed in the context of neurodiversity, and to identify organizational barriers that hinder the effective inclusion of neurodivergent developers.
\textit{Method:} We applied a mixed-method approach combining a web content analysis covering Reddit and LinkedIn with 11 semi-structured expert interviews from a corporate neurodiversity network in a German organization.
\textit{Results:} The analysis shows that teamwork practices are highly fragmented and shaped by individual adaptation rather than a shared standard. While agile practices and supportive tools can enable neurodivergent participation, rigid structures, stereotypes, and one-size-fits-all approaches often undermine inclusion. Organizational awareness and tailored adjustments remain insufficient.
\textit{Conclusion:} Agile practices can promote inclusive teamwork, yet their benefits are constrained by rigid organizational structures and limited awareness of neurodiversity. Harnessing neurodiverse strengths demands flexible organizational conditions and tailored support.

\end{abstract}
\section{Introduction}
Today, workplace diversity extends beyond demographics to include neurological differences, classified in ICD-11 as \textit{Neurodevelopmental Disorders} \cite{ICD_11_EN} and commonly described under the umbrella of neurodiversity \cite{Doyle.2020}.
Neurodiversity denotes natural variations in neurological development and cognitive processing~\cite{Rollnik.2024,ICD_11_EN}. Manifestations include autism spectrum disorder, attention deficit hyperactivity disorder (ADHD), dyslexia, or Asperger’s syndrome~\cite{Rollnik.2024,ICD_11_EN}. Today, accessibility is understood as the \textit{"result of environmental and societal barriers rather
than a characteristic of an individual"}~\cite{Rauschenberger_2024_DecadeAccessibility}.

In practice, organizations are increasingly acknowledging the potential of neurodivergent employees~\cite{Kersten.2025} and acknowledging neurodiversity~\cite{Burtscher.2024}. This means recognizing that cognitive differences can be a strength when supported by an inclusive work environment. However, organizational structures and processes that explicitly foster their strengths and facilitate their successful integration into teams remain rare~\cite{Das.2021} and remain insufficiently investigated~\cite{Rauschenberger_2024_DecadeAccessibility}. Employees face specific challenges, such as struggling to realize their perceived potential at work~\cite{Fuermaier.2021}. To increase diversity, Kersten et al.~\cite{Kersten.2025} emphasize the need for explicit management training at the organizational level and individualized work design to implement a resource-based and strength-based approach to inclusion.

Within the field of software development, agile methods have become widely established~\cite{VersionOne.2023}. These approaches rely on teamwork, self-organization, and continuous communication~\cite{Stray.2025}, principles that can present challenges to neurodivergent employees~\cite{Gama.2023}. At the same time, their distinct cognitive strengths, such as increased attention to detail and strong concentration, can constitute valuable assets for software development teams, particularly given the growing shortage of skilled professionals~\cite{Costello.2021,Morris.2015}. Nevertheless, neuroatypical employees often encounter structural barriers, are subject to misunderstandings, and are not adequately included in team processes~\cite{Gama.2025,Menezes.2025,daRocha.2024}.

Despite emerging corporate initiatives and an increasing interest in neurodiversity, empirical research remains limited~\cite{Rauschenberger_2024_DecadeAccessibility}. In particular, there is little systematic evidence on how agile teams should be structured and what constitutes an optimal team size to ensure the efficient and equitable inclusion of neurodivergent employees~\cite{Gama.2023,Morris.2015}. Moreover, organizational barriers that impede effective inclusion are still not sufficiently understood.

The above motivates our two research questions:
\begin{itemize}
    \item\textbf{RQ 1:} Which practitioner-reported teamwork practices, communication patterns, and adaptation strategies are discussed in public online discourse on neurodiversity in agile software development?\\
    In this study, we use \textit{teamwork quality} as a practice-oriented umbrella term covering collaboration-related aspects such as communication, coordination, team norms, perceived inclusion, and needs-oriented participation. We use the term in this broader exploratory sense to capture practitioner discourse on how agile teamwork is experienced and adapted in neurodiverse settings.
To address RQ1, we conducted an exploratory web content analysis (WCA) of public discussions on Reddit and LinkedIn.
    \item\textbf{RQ 2:} What organizational barriers prevent neurodiversity from being included in agile software development teams?\\
    To get an in-depth understanding of existing barriers for neurodiversity inclusion, we conducted a case study at a German company and applied in total 11 semi-structured interviews.     
\end{itemize}

This paper makes two contributions. First, based on an exploratory WCA, we synthesize practitioner discourse into four clusters that describe how agile teamwork is currently discussed with regard to neurodiversity, including communication patterns, task allocation, supportive tools, and process adaptations. Second, based on 11 semi-structured interviews in an agile organizational context, we identify nine organizational barriers that hinder the inclusion of neurodivergent employees in agile software teams. For practice, these findings translate into concrete adaptation points for communication, meeting formats, task allocation, and performance expectations. For research, the study refines the understanding of inclusion barriers in agile teamwork and distinguishes between general inclusion challenges and barriers that were described as specifically relevant in neurodiverse settings.

The paper at hand is structured as follows: First, we briefly explain the background of neurodiversity and give an outline of the related work in Section~\ref{sec:RelWork}. Next, we explain our mixed-method research design in Section~\ref{sec:ResearchDesign}. In Section~\ref{sec:Results} we present our results and answer the research questions. Before we close the paper with a conclusion in Section~\ref{sec:Conclusion}, we discuss the practical limitations in Section~\ref{sec:PracticalImplications}. 

\section{Background \& Related Work}
\label{sec:RelWork}
In this Section, we first give a brief background of neurodiversity aiming to provide a ground of fundamental knowledge for this paper. Next, we present an overview of the related work with the objective to synthesize existing research findings and point to the new contributions of our study to the body of knowledge in agile software development.

\subsection{Background on Neurodiversity}
Neurodiversity refers to the natural variation in neurological development and cognitive functioning that exists across the human population. This concept includes conditions listed in the ICD-11 \cite{ICD_11_EN} as neurodevelopmental disorders. Examples include autism spectrum disorder, attention deficit hyperactivity disorder (ADHD), dyslexia, and Asperger’s syndrome. It understands these conditions as part of normal human variation rather than as deficits. Current accessibility research argues that many difficulties faced by neurodivergent individuals stem from environmental and societal barriers. Such barriers include communication demands, literacy requirements, and norms related to attention and behavior~\cite{Doyle.2020,Khan.2023,LeFevre.2023}. They are therefore not located within the individual. The neurodiversity perspective highlights the strengths associated with different cognitive profiles. It also stresses the need for inclusive environments that accommodate these differences. The focus thus shifts from remediation to recognition, empowerment, and equitable participation to uplift the person and organization.

\subsection{Related Work}
In the area of agile software development, the research landscape remains somewhat limited. We thus decided to search for related literature with a broader context scope and tried to identify studies dealing with neurodiversity in the field of software engineering. We searched for primary and secondary studies using Google Scholar. Below, we provide a brief overview of the work closely related to our study.

Rather than framing the diverse manifestations of neurodiversity as cognitive deficits, we should emphasize their potential strengths~\cite{Gama.2023}. 
In software development contexts such as programming, testing, and quality assurance, traits like enhanced pattern recognition, attention to detail, and sustained concentration can be particularly valuable~\cite{Gama.2023}. 
Harnessing these abilities may yield productivity gains, foster innovation, and enrich team diversity~\cite{Liebel.2024}. 

Nevertheless, research by other authors (\textit{e.g.}, ~\cite{Gama.2025,Liebel.2024}) highlights the challenges that developers with neurological conditions such as ADHD encounter in their daily work. Reported difficulties include insufficient consideration of individual communication needs, limited awareness of neurodiversity within teams, and problems related to time management and task prioritization~\cite{Liebel.2024}.

In their exploratory interview study, Morris et al.~\cite{Morris.2015} provide a systematic account of both the strengths and the structural barriers experienced by neurodivergent developers, thereby complementing the findings of Liebel et al.~\cite{Liebel.2024}. Identified barriers include sensory overload, as well as the fear of stigmatization associated with the disclosure of neuroatypical characteristics. The authors advocate for a more individualized approach to neurodiversity in technical work environments.

More recent studies support the conclusions of Morris et al.\cite{Morris.2015}, indicating that agile practices can pose challenges to neurodivergent team members~\cite{Gama.2023}. If applied in an unstructured way or without individual adaptation, such practices can become burdensome and negatively affect work outcomes~\cite{Gama.2023,Liebel.2024}.

Despite growing awareness of neurodiversity in organizations, research on specific barriers in working practices, communication processes, and adaptation strategies at the team level in software development, the research landscape remains lacking. Existing studies predominantly focus on the recruitment of neurodivergent employees or emphasize normative appeals for inclusion. Thus, our study addresses an important research gap that identifies specific barriers to integrate neurodivergent employees into agile teams.

\section{Research Design}
\label{sec:ResearchDesign}
To examine the integration of neurodiversity in agile software development teams, we employed a mixed-methods approach. Because empirical research in this area remains scarce, the study combined a Web Content Analysis (WCA) of discussions on Reddit and LinkedIn, as well as eleven semi-structured expert interviews drawn from a corporate neurodiversity network. This design made it possible to capture both the breadth of ongoing discourse and the depth of organizational experiences with inclusion. 

Methodological rigor and transparency were ensured through a detailed research protocol that provides more detailed information on how we applied the study, \textit{i.e.,} the data analysis artifacts like our coding book. The research protocol is available at Zenodo~\cite{Struck.2025}.

\subsection{Web Content Analysis}
To address the first research question, we conducted a WCA. This method was selected, because it enables the systematic exploration of current practices and discourse beyond the scope of peer-reviewed literature. A WCA enables real-time assessment of the state of practice~\cite{Garousi.2020} and is widely adopted in Software Engineering research~\cite{Bagheri.2016}. For our analysis, the online platforms \textit{Reddit} and \textit{LinkedIn} were selected: 
\textit{Reddit} is a widely used international forum where software developers, among others, engage in anonymous discussions on a broad range of topics. 
\textit{LinkedIn}, by contrast, represents the leading professional networking platform, facilitating the exchange of information between employees and employers in a professional context. The combination of these two sources was chosen to mitigate potential biases—such as the predominance of personal anecdotes on Reddit or the greater emphasis on topics related to inclusion on LinkedIn—thereby ensuring a more balanced perspective. The collection, analysis, and clustering of content followed the systematic methodological approach by Kim and Kuljis~\cite{Kuljis.2010}. 

In our WCA, an item refers to one public post or discussion thread returned by the platform search and considered as a unit of analysis. Reddit and LinkedIn were chosen deliberately because they represent two complementary discourse settings: Reddit allows comparatively candid and often experience-based discussion under pseudonymity, whereas LinkedIn reflects more professionalized and organization-facing discourse. We did not aim at representativeness across all platforms, but at contrasting two relevant practitioner spaces with different communication norms.

We iteratively tested several search strings on both platforms and retained those that yielded the highest proportion of relevant results. We then applied explicit inclusion and exclusion criteria. We included items that referred to neurodiversity or neurodivergence in work or software-related contexts and contained substantive discussion of teamwork, collaboration, communication, agile practices, or inclusion barriers. We excluded duplicates, job advertisements, purely promotional posts, short reactions without analyzable content, and posts not sufficiently related to agile teamwork or software work. The final search was conducted on 17 June 2025. This yielded 17 Reddit items and 35 LinkedIn items at screening stage; after applying the criteria, 4 Reddit items and 5 LinkedIn items remained for analysis. The complete search strings, decision rules, and the selection overview are available in the research protocol~\cite{Struck.2025}.

Relevant content from the selected posts was systematically extracted and organized using a Miro board. A predefined coding scheme, described in Section Interviews of the research protocol~\cite{Struck.2025}, guided the cluster-based thematic analysis. The material was categorized into overarching themes, with color-coded markers used to differentiate between sources (\textit{e.g.,} Reddit contributions were labeled with green post-it notes). This procedure enabled a transparent and systematic synthesis of practitioner perspectives on teamwork practices and neurodiversity in agile software development teams.

\subsection{Qualitative Case Study}
We conducted a single case study to examine organizational barriers to the inclusion of neurodivergent employees in agile software development teams. Building on the insights of the WCA, the case study provided a deeper understanding of barriers and contextual dynamics within an organizational setting. The case study was designed and conducted according to the guideline of Runeson and Höst~\cite{Runeson.2009}. Case studies are particularly suitable for investigating contemporary software engineering phenomena in their natural environment, especially when the boundaries between the phenomenon and its context are not clearly defined~\cite{Yin.2009}. The study was carried out between April and June 2025, with participants recruited from a corporate neurodiversity network in a German organization.

\textbf{Case Study Context}
The single case study was conducted at Hogwarts Express (anonymized), a German organization operating in the railway industry. The company employs more than 7,000 people and operates in three locations in Germany (Frankfurt/Main, Berlin, and Erfurt). More specifically, the case study was situated within a defined organizational context. Hogwarts Express established a diversity network seven years ago, dedicated to fostering the inclusion of neurodivergent employees. Within this network, employees from various divisions of the organization collaborate to promote resilient and sustainable approaches to inclusion. Activities include the facilitation of Communities of Practice as well as company-wide initiatives such as \textit{Diversity Week}, which the network plays a central role in organizing and implementing. For the purpose of this case study, we deliberately selected interview participants from the diversity network of Hogwarts Express who were directly connected to agile software development.

\textbf{Qualitative Data Collection}
Our qualitative data collection comprises 11 semi-structured interviews designed to ensure comprehensive coverage of the phenomena under investigation. The selection of interviewees reflects a cross-section of organizational levels, roles, and departments. As an inclusion criterion, participants needed experience with agile methods and a concrete connection to neurodiversity in their work context. All interviewees were active in the company’s neurodiversity network and were selected because they could reflect on neurodiversity in relation to agile collaboration from their organizational role and experience. However, the study did not differentiate systematically between self-experience, observation, support, or leadership-related experience with neurodiversity. Table~\ref{tab:participant-overview} presents an overview of the participants, including their current roles and experience with agile practices.

\begin{table}[ht]
\centering
\begin{tabular}{@{}>{\centering\arraybackslash}p{1cm}|>{\centering\arraybackslash}p{3.5cm}|>{\centering\arraybackslash}p{5cm}@{}}
\textbf{ID} & \textbf{Role} & \textbf{Exp. with Agile Methods} \\ 
\hline
P01         & Team Lead               & 14 years                        \\
\hline
P02         & Product Owner             & 8 years                     \\
\hline
P03         & Scrum Master             & 20 years                     \\
\hline
P04         & Scrum Master           & 15 years                      \\
\hline
P05         & Team Member                                 & 5 years  \\
\hline
P06         & Scrum Master                             & 10 years  \\ 
\hline
P07         & Team Member                           & 3 years  \\ 
\hline
P08         & Team Member                           & 8 years  \\  
\hline
P09         & Team Member                         & 10 years  \\
\hline
P10         & Scrum Master                         & 7 years  \\
\hline
P11         & Scrum Master                         & 26 years  \\  
\hline
\end{tabular}
\vspace{0.5em}
\caption{Interviewee profiles}
\label{tab:participant-overview}
\end{table}

The interviews followed a predefined guideline, which is available in our research protocol (see Appendix C in \cite{Struck.2025}). The guideline comprises 35 questions in total: 8 questions to capture the interviewees’ background, 8 questions addressing team culture and quality, 12 questions focusing on agile practices applied within the respective teams, and 7 questions concerning the influence of neurodiversity and the inclusion of neurodivergent individuals.

The interviews were conducted virtually via Microsoft Teams in May 2025 and were held in German. All 11 participants provided informed consent for audio recording. Each interview lasted between 33 and 70 minutes. Two researchers facilitated the process: one conducted the interview, while the other managed technical aspects, including recording and taking notes. The raw interview data is not available for the public due to confidential reasons.

\textbf{Qualitative Data Extraction \& Analysis}
For qualitative data analysis, recorded interviews from the Hogwarts Express case were transcribed using \href{https://voice.ai/}{\texttt{VoiceAI}}. The transcripts were subsequently reviewed and refined by the first four authors to ensure both accuracy and completeness. These validated transcripts provided the basis for a thematic analysis, a method used to identify, analyze, and report patterns in qualitative data. The analysis followed the framework proposed by Braun and Clarke~\cite{Braun_Clarke.2006}, consisting of five phases and subsequent reporting: data familiarization, initial code generation, theme identification, theme review, and theme definition and naming.

Following the above mentioned approach by Braun and Clarke~\cite{Braun_Clarke.2006} and to strengthen the reliability and consistency of the coding process, the first four authors coded a subset of the data in pairs and compared their results with the coding produced by the last author. Any discrepancies were addressed through collaborative discussions, fostering a shared understanding of the data and enhancing the robustness of the thematic framework. This iterative refinement ensured that the themes were firmly grounded in the data while remaining aligned with the study’s objectives.

\begin{figure*}[t]
\centering
\includegraphics[width=\textwidth]{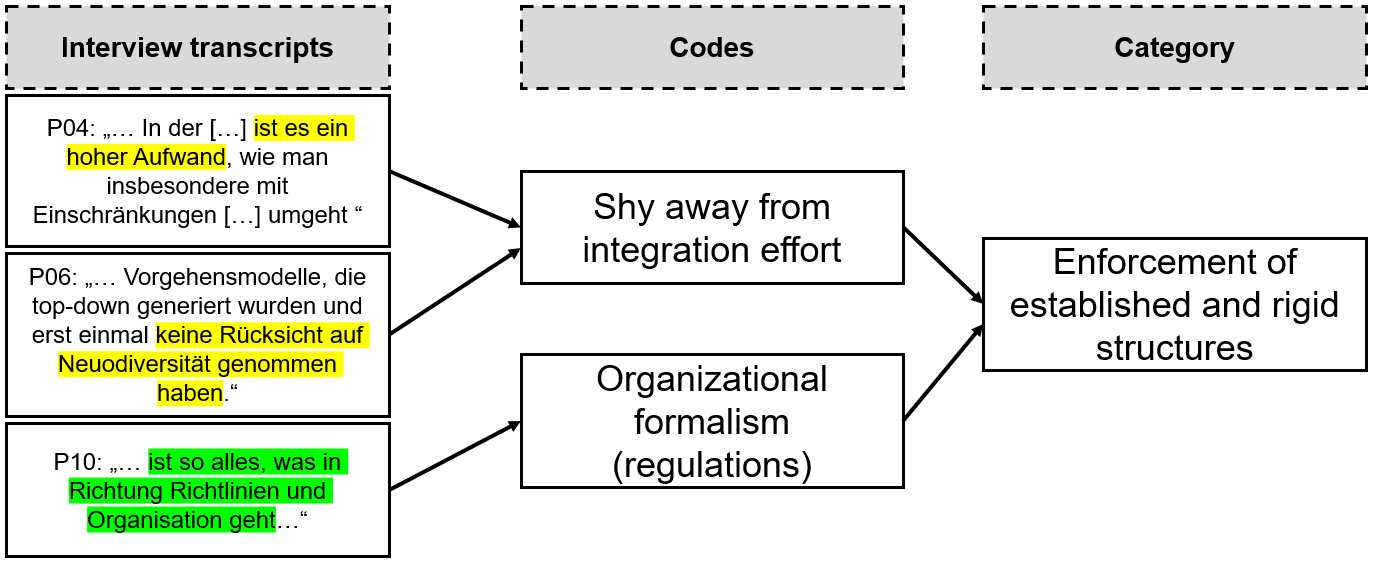}
\caption{Coding and Categorization Example}
\label{fig:CodingExample}
\end{figure*}
The thematic analysis yielded nine clusters of barriers, which were documented and visualized on a Miro board. Figures 2 and 3 in our research protocol~\cite{Struck.2025} depict graphical excerpts of this board, while Figure~\ref{fig:CodingExample} illustrates a specific example of the coding and clustering process.

\subsection{Threats to Validity}
Like any empirical investigation, our study is subject to certain limitations that arise from the chosen research design. While the study was carried out in line with the relevant guidelines, we acknowledge these limitations and outline the measures taken to reduce their potential impact. For this purpose, we applied the \textit{Threats to Validity} framework as proposed by Wohlin et al.~\cite{Wohlin.2012} and Runeson and Höst~\cite{Runeson.2009}.

\textbf{Construct validity:} Related to the applied WCA, we especially address the quality of the data sources used for construct validity. It is worth mentioning that posts on Reddit and LinkedIn may be shared or reviewed without much scrutiny or regulations for publication~\cite{Proferes.2021}. 

\textbf{Internal validity:}  We implemented specific measures to ensure consistency in interviewing and to reduce potential bias. First, the interview guideline consisted of neutral, non-leading questions. Moreover, the semi-structured design allowed us to explore relevant topics in greater depth, depending on the interviewees’ responses. To enhance rigor, each interview was facilitated by at least two researchers. Nevertheless, since the study was conducted within a single company, contextual factors such as organizational culture or external regulations may have shaped the perceived barriers and benefits.

\textbf{External validity:} The interview results are based on the experiences of eleven Hogwarts Express employees. Although participants from various roles and departments were included, it is plausible that this composition influenced the outcomes. Furthermore, the findings are closely tied to the specific case context and therefore may not be fully transferable to other industries, organizational sizes, or cultural settings.

\section{Results}
\label{sec:Results}
In this Section, we present the results of our study and answer our two research questions below in the following subsections. 

\subsection{State of the Practice of Teamwork Quality in Agile Teams}
Here, we answer our first research question: \textit{What is the current state of practice regarding teamwork quality in agile software development teams in the context of neurodiversity inclusion?}
Our WCA revealed a wide range of recommendations and personal experiences. Our analysis indicates that it is not possible to identify a single, definitive state of practice regarding the quality of teamwork in agile software development. The topic is characterized by considerable breadth and complexity, and remains underexplored in empirical research in software engineering, which hinders the formulation of conclusive statements. To capture the current discourse, we therefore synthesized the material into four overarching thematic clusters.

\textbf{Applied Software Processes:} Agile practices offer several advantages for neurodivergent team members. Daily stand-ups anchor accountability, retrospectives create space for problem identification, and short iterations prevent distraction. QA tools and automated tests help to follow standards, while asynchronous stand-up updates allow introverted or colleagues with language delays to participate at their own pace. These methods structure the workflow, provide regular synchronization points, and break tasks into manageable units, especially beneficial for those with ADHD or autism by strengthening their focus and motivation. However, some warn that rigidly applying these rituals without individual adjustment can backfire and negate their benefits.

One challenge of remote work is that non-onsite work organization reduce nonverbal communication, which can pose difficulties even for neurotypical colleagues. An increasingly remote world can potentially hinder effective team collaboration. To counteract this, deliberately including more neurodivergent members in distributed teams can help, since they often rely less on nonverbal signals. Their different processing styles tend to resist unconscious conformity and frequently excel at clear written and asynchronous communication.

\textbf{Individual Task Allocation:} The tasks should be deliberately matched to the cognitive strengths and interests of each neurodivergent member, such as logic, pattern recognition, or organization, to increase engagement. Providing autonomy to choose tasks can help avoid undesired and unnecessary administrative work. When the fit is poor, productivity can quickly drop. Management roles can also be appealing if they allow room for innovation, experimentation, and flexibility. Conversely, rigid reporting and planning routines tend to be demotivating.

\textbf{Supportive Tools:} Employee Resource Groups (and sub-groups specifically for neurodivergent employees) create safe spaces for exchange, shared experiences, and joint events. Mentoring by industry colleagues with similar challenges, for example experienced developers with ADHD, offers practical support. Some organizations use communication guides during onboarding, where individuals outline their strengths, preferred communication styles, and desired forms of support. This way, teams know from day one how to collaborate most effectively.

\textbf{Communication:} Communication becomes a major challenge. Many neurodivergent employees feel overwhelmed and bored in technical discussions, and they lose track of both content and pace when taking notes. Recommended strategies include requesting recording and transcription permissions to review conversations later or faster and sending follow-up emails that confirm agreed-upon tasks in writing. Visual aids such as diagrams, brief check-ins after each segment, and asynchronous stand-up updates, whereby team members post their updates in a shared chat before the meeting, are also considered. This allows those who prefer written communication to contribute in a way that suits them.
Feedback in the workplace can be either motivating or demotivating. Neurodivergent employees are especially vulnerable when feedback is based on subjective impressions or when it labels personality traits as deficiencies. Neuroinclusive feedback processes avoid these distortions by following a three-step sequence:

\begin{enumerate}
    \item contextualize the observation,
    \item describe in a precise way its impact on the team and tasks,
    \item collaboratively plan concrete support measures
\end{enumerate}

\subsection{Neurodiversity Inclusion Barriers}
Based on the results of the interviews, we answer our second research question: \textit{What organizational barriers prevent neurodiversity from being included in agile software development teams?}

\begin{figure*}[t]
\centering
\includegraphics[width=\textwidth]{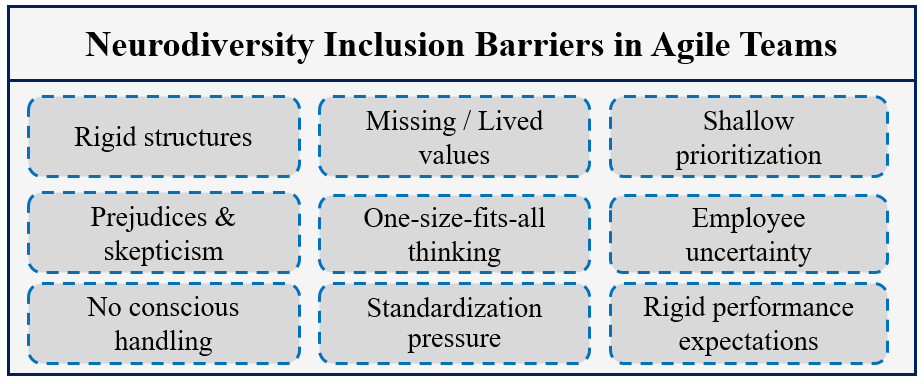}
\caption{Overview of the identified neurodiversity barriers}
\label{fig:Results}
\end{figure*}

The inclusion of neurodivergent software developers can be hindered by various barriers that need to be overcome to foster inclusion in organizations. These barriers can affect the productivity, integration, and well-being of those who seek to be included. 
Based on the conducted interviews, we identified various barriers and categorized them into nine different clusters, which we explain below. The identified barriers represent the major ones during the inclusion of neurodivergent developers in agile teams, based on the conducted interviews. Figure~\ref{fig:Results} depicts the identified barriers. 

\textbf{Enforcement of established and rigid structures:} 
According to the interviewees, the strongest barrier to the inclusion of neurodivergent developers lies in rigid organizational structures. These structures manifest in unadapted frameworks, inflexible performance requirements, and administrative processes that leave little room for individual needs. As a result, developers may experience performance pressure, reduced motivation, and overwhelming stress. \textit{Interviewee 2} emphasized that managers often attempt to work around these constraints, but their options remain limited: “[...] \textit{thinking upward and outward, the instruments and regulations of Hogwarts Express still come from another time, and they are not always suitable.} [...] \textit{But when corporate headquarters or HR departments determine that next year things must be done according to a specific catalog, that catalog often fails to meet the needs of neurodivergent employees. The situation becomes even more precarious when it concerns leadership roles.}”

\textbf{Prejudices and skepticism:} Several interviewees named prejudices and skepticism of coworkers and managers as a major barrier affecting inclusion. Missing acceptance, accordance, uncertainty and insufficient awareness can create an unstable environment for neurodivergent developers. These barriers can be caused by skepticism and missing awareness. According to interviewee 4 a difference between generations can be observed. Especially older generations are less tolerant and understanding. This skepticism can cause divisions within a team: "[...] \textit{among older people, for example, ADHD is very clearly coded as hyperactive, screaming children who run around, tear things apart, and break everything. And when you then try to apply that to adults, people simply do not take you seriously.} [...] \textit{Some even outright deny the existence of neurodivergence or say things like, ‘Well, if you have ADHD, then I must have it too. Don’t we all have a bit of ADHD?’ And that kind of dismissal is something many people do not realize is actually hurtful. These awareness workshops are meaningful in that sense, but many people then feel patronized or respond with rejection.}"

\textbf{No conscious handling of neurodivergent developers:} No conscious handling of neurodivergent developers describes neurotypical behavior towards neurodivergent employees, who seek to be included. When neurodivergent developers are treated, evaluated and seen as neurotypical developers, the needed inclusion is not present. As a result of this behavior, inclusion measures are either not implemented or not prioritized, as the underlying need is overlooked. This barrier is not caused by malicious intent, but rather by lack of awareness or knowledge.

\textbf{Fitting developers to a specific standard:} Based on the interviewees, neurodivergent developers often get standardized into a specific expectation, which cannot be met. Unrealistic and excessive demands and expectations can cause disappointment within teams, as well as higher organizational levels. According to interviewee 4 neurodivergent developers often need more time to adapt to a new team. Once adapted, a deeper thematic understanding can be observed. Not granting this time leads to developers not meeting these expectations and possibly getting penalized by the organization. Interviewee 4 stated: "[...] \textit{because I simply did not perform. That was because, in parallel, I was — as I personally describe it — putting down roots: I reach out left and right into all kinds of fields at the same time, but then I try to gradually go deeper. And once I have managed that, I become the specialist expert.}"

\textbf{Not living team values and missing consensus:} A unified set of values is a necessary requirement for a positive and good work environment inside and outside development teams. A missing value set, lack of a consensus, or not living those values can cause a divide within a team. Identical team and organization values are the foundation for a positive environment. Values like tolerance, diversity, and equality need to be actively lived to be effective. Even though organizations often communicate these values, there can be a lack in the implementation. Interviewee 10 mentioned: "\textit{What I perceive from top management is that diversity holds an important position — in the sense of acceptance. But I think what is still somewhat lacking is that this is truly lived out within the teams. I increasingly see many teams that are not very diverse. Even just in terms of gender distribution, there is definitely room for improvement. I believe that acceptance and openness are present, but the actual practice — the real implementation — is not yet as pronounced as I would wish.}"

\textbf{Using one-size-fits-all solutions:} Interviewees described uniform ways of working as a recurrent barrier when teams expected all members to participate in the same formats, rhythms, and communication modes. This became visible, for example, when attendance in synchronous meetings was treated as the default, when retrospectives relied on spontaneous verbal reflection only, or when all team members were expected to work equally well in open office environments. In such situations, the problem was not agility as such, but the rigid application of one collaboration format to all members. Interviewees therefore emphasized the need for local adaptation, including flexible workplace choices, written input before meetings, asynchronous status updates, and meeting formats that reduce time pressure and spontaneous turn-taking.

\textbf{Shallow prioritization:} When choosing to include neurodiverse team members, these inclusions need to be supported steadily. External factors can reduce this support, which endangers the strength and success of inclusion. According to interviewee 3 a system under pressure often works differently: "\textit{A lot of work has already been done there, I would say. My concern is with maintaining it, because, as you know, the railway} [in Germany] \textit{has come under considerable pressure. And systems under pressure generally behave differently from what was planned over many years. This means one has to see whether things get watered down, or whether the level of inclusion and diversity remains as it is at the moment. It is certainly desirable. But I do not believe it can be cemented, because in the end it depends on the people working together.}"
Especially during financial crisis, these dangers can occur. Based on the interviewees, the pressure to save money can be a major problem when deciding on the priority of neurodivergent inclusions. While in crisis an organization might reduce the level of diversity and inclusion. Assistance for neurodivergent developers might decline during these times.

\textbf{Wrong handling through uncertainty:} The handling of neurodivergent developers can be an initial barrier for neurotypical developers without experience in inclusion. Developers without prior experience with neurodiversity can become overwhelmed by their own uncertainty. Based on interviewee 6 the intention not to handle neurodiversity wrong can cause neurotypical developers not to act: "\textit{When something is relevant for the group or within this group dynamic, it does not necessarily mean that it is also relevant for the individual need. Raising awareness and creating understanding in these areas can be a long journey. In an agile world, we often operate at a high frequency, in very short coordination cycles. We do not always have the space and time to address these issues, even though they would require it and, from my perspective, would improve collaboration.}" Therefore uncertainty and inexperience with neurodiversity represent a barrier to inclusion.

\textbf{Non-adapted performance requirements:} Based on their neurodiversity, developers might need adapted performance requirements. Lack of adaptation can represent a barrier for the inclusion measures. According to interviewee 10 these developers often work slower but more thoroughly: "\textit{The acceptance that some people — I would not say that they are slower, but perhaps because they think a little more out of the box — may want to improve things in the context of the topics they are working on. In other words, they want to create sustainable added value for the team, the project, and the vision. But then, to put it somewhat casually, they may end up doing a bit more together than was originally defined. And this can lead to conflicts.}"
This potential cannot be extracted without adapting the expectations and performance requirements. Too high expectations or too fast judgment affects the developer and the team, because work might get redistributed or developers might get penalized. Interviewees did not describe lower capability, but a mismatch between standardized performance expectations and different working rhythms. In particular, they reported tension when speed was valued more strongly than depth, when exploratory problem solving was penalized as inefficiency, or when adaptation time in new teams was not granted.

\section{Discussion}
\label{sec:PracticalImplications}
Our findings confirm prior work showing that agile practices can both support and burden neurodivergent employees, depending on how rigidly they are applied. At the same time, our study extends previous work in three ways. First, it shifts the focus from individual coping to organizational and team-level barriers. Second, it shows that many inclusion problems emerge not from agile practices per se, but from their standardized implementation. Third, it identifies concrete adaptation points in meeting design, communication, task allocation, and performance evaluation.

Some barriers identified in our data are not exclusive to neurodivergent employees. For example, unclear team values, shallow prioritization of inclusion, and rigid organizational structures can also affect newcomers or other marginalized groups. However, interviewees described these barriers as becoming particularly salient in neurodiverse settings when teams relied heavily on spontaneous verbal interaction, implicit norms, and standardized performance expectations. We therefore interpret our findings as partly general team-level inclusion problems and partly as intensified or differently manifested barriers in neurodiverse agile contexts.

Our findings suggest that inclusive agility is less about replacing agile methods and more about introducing configurable options within them. Examples include asynchronous stand-up updates, written retrospective input before group discussion, visual meeting agendas, explicit communication agreements, flexible task matching, and evaluation criteria that acknowledge differences in working rhythm and depth of processing.

\section{Conclusion \& Future Work}
\label{sec:Conclusion}
This study examined the integration of neurodiversity in agile software development teams using a mixed-methods approach. We first assessed current practices in team culture, communication, and needs-based inclusion, and then identified organizational barriers to the participation of neurodivergent employees. Findings show that agile rituals can offer helpful structure but risk overregulation when applied too rigidly, while practices such as asynchronous updates, visual aids, and individualized communication support focus and motivation. Employee resource groups and mentoring programs further help raise awareness and close knowledge gaps.

The main barriers identified are rigid organizational structures, persistent prejudices, and the difficulty of addressing individual needs without one-size-fits-all solutions. These challenges are compounded by unadjusted performance expectations during high-stress periods and by uncertainty about how to respond to individual requirements. Rather than suggesting a new agile framework, our findings indicate that existing agile practices should be applied with greater configurability. In our data, this concerned especially communication channels, meeting formats, task allocation, and performance expectations. Examples mentioned by participants included asynchronous updates, written feedback options, visual aids, and more individualized collaboration agreements. In addition, continuous awareness-building and targeted training for leaders and staff are crucial to institutionalize inclusive practices. To assess the effectiveness of these measures, future research should employ quantitative longitudinal studies and compare different agile frameworks. Overall, our study suggests that the main challenge is not the presence of agile rituals themselves, but the assumption that one uniform way of enacting them fits all team members. Future work should therefore examine which concrete adaptations are effective for which neurodivergent profiles and organizational settings, and where barriers are specific to neurodiversity rather than general team inclusion.

\section*{Acknowledgements \& Final Remark}
We would like to express our gratitude to the case company for supporting this study and to all interviewees for their participation. As a final note, we acknowledge that several sections of this paper were refined with the assistance of the GenAI tool ChatGPT (GPT-5.0).

\bibliographystyle{splncs04}
\bibliography{references}

\end{document}

%% file: aux-commands.tex
\usepackage{ifthen}

\newboolean{auxinfo}
\setboolean{auxinfo}{true}   

\newboolean{showoutline}
\setboolean{showoutline}{true}

\newboolean{showcomment}

\setboolean{showcomment}{true} 

\usepackage[subtle]{savetrees}

\ifthenelse{\boolean{auxinfo}}
  {}
  {
    \setboolean{showoutline}{false}
    \setboolean{showcomment}{false}
  }

\usepackage{xcolor}
\usepackage{amssymb}

\ifthenelse{\boolean{showcomment}}
   {\newcommand{\nb}[2]{\fcolorbox{gray}{yellow}{\bfseries\sffamily\scriptsize#1}{\sf\small$\blacktriangleright${\em #2}$\blacktriangleleft$}}
   \newcommand{\working}[1]{\fcolorbox{gray}{yellow}{{\bf #1}\emph{\scriptsize---in progress---}}}
   \newcommand{\TBD}[1]{\fcolorbox{gray}{yellow}{{\bf #1}\textbf{TBD}}} 
  }
  {\newcommand{\nb}[2]{}{}
   \newcommand{\working}[1]{}
   \newcommand{\TBD}[1]{} 
  }

\usepackage{todonotes}
\ifthenelse{\boolean{showoutline}}{
	\newcommand{\outline}[3]{
		~\newline 
		\fcolorbox{red}{white}{
			\parbox{\columnwidth}{
				\ifthenelse{\equal{#1}{}}{
					\ifthenelse{\equal{#2}{}}{
						\noindent\colorbox[rgb]{0.65,0.16,0}{\textcolor[rgb]{1,1,1}{\textbf{Outline}}}
					}{
						\colorbox[rgb]{0.65,0.16,0}{\textcolor[rgb]{1,1,1}{\textbf{Outline -- Responsible: #2}}}
					}
				}{
					\ifthenelse{\equal{#2}{}}{
						\noindent\colorbox[rgb]{0.65,0.16,0}{\textcolor[rgb]{1,1,1}{\textbf{#1 page(s)}}}
					}{
						\colorbox[rgb]{0.65,0.16,0}{\textcolor[rgb]{1,1,1}{\textbf{#1 page(s) -- Responsible: #2}}}
					}
				}
				#3
			}
		}
	}
}{
	\newcommand{\outline}[3]{}
}

\newcommand\defauxcomm[1]{
       \expandafter\newcommand\csname #1\endcsname[1]{\nb{#1}{##1}}
       \expandafter\newcommand\csname WK#1\endcsname{\working{#1}}
       \expandafter\newcommand\csname TBD#1\endcsname{\nb{#1}}
    } 
   
\defauxcomm{KS}
\defauxcomm{MN}

\usepackage[normalem]{ulem}
\ifthenelse{\boolean{showcomment}}
    {\newcommand{\strike}[1]{\textcolor{red}{\sout{#1}}}}
    {\newcommand{\strike}[1]{}}